# Triggering one dimensional phase transition with defects at the graphene zigzag edge

Qingming Deng[1] and Jiong Zhao[1,2,3*]

1, IFW Dresden, Institute of Solid State Research, P.O. Box 270116, D-01171 Dresden, Germany 2, Center for Integrated Nanostructure Physics, Institute for Basic Science(IBS), Sungkyunkwan University, Suwon 440-746. Korea 3, Department of Energy Science, Department of Physics, Sungkyunkwan University, Suwon 440-746, Republic of Korea

**ABSTRACT:** One well-known argument about one dimensional(1D) system is that 1D phase transition at finite temperature cannot exist, despite this concept depends on conditions such as range of interaction, external fields and periodicity. Therefore 1D systems usually have random fluctuations with intrinsic domain walls arising which naturally bring disorder during transition. Herein we introduce a real 1D system in which artificially created defects can induce a well-defined 1D phase transition. The dynamics of structural reconstructions at graphene zigzag edges are examined by in situ aberration corrected transmission electron microscopy (ACTEM). Combined with an in-depth analysis by ab-initio simulations and quantum chemical molecular dynamics (QM/MD), the complete defect induced 1D phase transition dynamics at graphene zigzag edge is clearly demonstrated and understood on the atomic scale. Further, following this phase transition scheme, graphene nanoribbons (GNR) with different edge symmetries can be fabricated, and according to our electronic structure and quantum transport calculations, a metal-insulator-semiconductor transition for ultrathin GNRs is proposed.

## Introduction

It's widely accepted that one dimensional(1D) system with short-range interaction cannot have 1D phase transition at finite temperature (above zero).[1] The statistical random fluctuations caused by entropy contribution and non-interactive nature between domain walls make 1D phase transition rare. However as an example on the graphene (monolayer $sp^2$ carbon) edge we will introduce, defects in lattice can play a crucial role to trigger 1D phase transition.

The edge states of graphene have been emphasized a lot for their abilities of tuning band gap[2,3] magnetic moment[4,5] or functionalization with other dopants.[6,7] Basically there are two types of most stable graphene edges, zigzag (ZZ) and armchair (AC),[8] but due to the edge passivation and edge stress[9,10] various edge reconstructions might take place, including ZZ(57), AC(677), AC(57), etc.,[6,10,11] the number in brackets refers to the number of member carbon atoms in each ring in one period at graphene edges, i.e. ZZ(57) means the zigzag edge reconstructed with alternating pentagon-heptagon structures.[12] Some theoretical calculations (taking into account spin polarization) showed the thermodynamic stability of the edges: ZZ(57) > AC > AC(677) > ZZ[13,14] while direct transmission electron microscopy (TEM) observations also demonstrated the reconstructed ZZ(57) edges, with some effect of electron beam.[15-19] However, the complete phase transition dynamics, the role of the electron beam (in TEM) and the effect of the unit cell size change (reconstructed ZZ(57) edge with twice the unit cell length along edge direction than the original ZZ(66) edge) are still not clarified yet.

Nucleation processes are required for phase transitions.[20] With the slightly higher thermodynamic stability (ground state) for ZZ(57) edge than the original ZZ edge evidenced by density functional theory (DFT),[13,14] some molecular dynamics (MD) calculations have been carried out to show that the first unit cell transition (nucleation) for ZZ(66)-ZZ(57) has an energy barrier 0.83eV at room temperature, and decreases to 0.5eV in the following second unit cell transition.[11] The higher barrier in the first cell transition is caused by the local strain mismatched with neighboring unit cells. Under room temperature TEM measurements, required energy to overcome this barrier is possibly supplied by the thermal energy, as well as by the high energy electron irradiation effect. Because of the good electrical conductivity and thermal conductivity, knock-on effect is the major part for graphene damage by the TEM. The displacement threshold ($T_d$) for a carbon atom on ZZ(57) edges is ca. 16 eV while for ZZ(66) edge it is ca. 14 eV,[13] implying slightly higher dynamic(under irradiation) stability for ZZ(57) than ZZ(66). Actually both the ZZ(57) and ZZ(66) edges have the Td very close to the knock-on damage threshold (16 eV) for $sp^2$ carbon under 80 kV electron beam.[21]

In this study, we employed in situ high resolution TEM (HRTEM) to unveil the reconstruction processes on graphene ZZ edges. The initialization of the reconstruction can be triggered by an electron beam knock-on defect, followed by a 1D phase transition over the whole edge. In particular, the two states ZZ(57) and ZZ(75) are reversibly transformed, which opens more opportunities to control the symmetry of graphene nanoribbons(GNR). Our quantum transport calculations present different electrical transport behaviors for these ultrathin ribbons with different symmetries.

## Results and Discussion

The graphene sample is prepared by mechanical cleavage method,[22,23] and then transferred onto conventional copper TEM grid. An imaging aberration corrected TEM working under 80kV[23] was similar to the previous studies.[24,25] Various defects or holes can be created by electron beam in monolayer graphene, which made it possible to find different kinds of freshly prepared graphene edges (ZZ or AC) along the boundaries of big holes, nevertheless we focus on ZZ edges in current work.

Figure 1 presents the optimized atomic models by DFT calculation (Figure 1c,d)[23] and the corresponding HRTEM images (Figure 1a,b), as well as the simulated images from the relaxed models with a multislice method (Figure 1c,d),[23] for the ZZ(66) edge and ZZ(57) edge, respectively. The unit cell length of ZZ(57) is twice of ZZ(66). Although in view of symmetry, ZZ(57) and ZZ(75)(defined from left to right, shown in the inset of Figure 1b) edges are exactly the same, however after reconstruction the ZZ(57) edge loses one inversion symmetry compared to ZZ(66), especially the distinction between ZZ(57) and ZZ(75) edges lies in the formation of nanoribbons with two parallel ZZ edges. Further, except for the ZZ(57) edges, some other defects on ZZ(66) and ZZ(57) edges, can also be found and identified in the TEM image series.

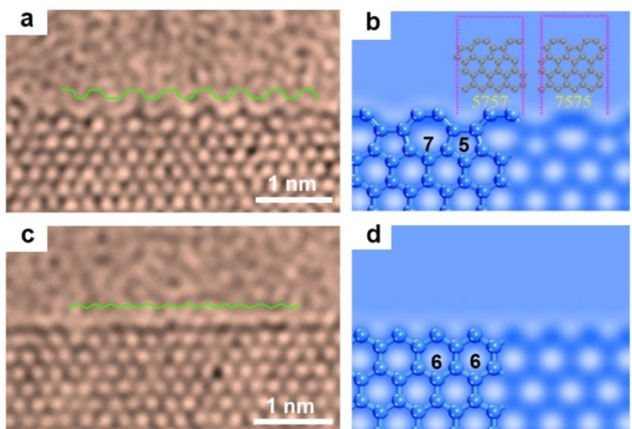

**Figure 1** (a)TEM image for the reconstructed graphene zigzag(57) edge. (b) Simulated TEM image overlaid with DFT calculated atomic structures for 57 edge. Inset presents the definitions of 57 type and 75 type edges. Atoms in pink rectangle are fixed during the simulations. (c) TEM image for the zigzag(66) edge. (d) Simulated TEM image overlaid with DFT calculated structures for 66 edge.

The bistablility of ZZ(57) and ZZ(66) edges is confirmed by the TEM image series which represent some stationary states(Figure 2). Moreover, continuous exposure to the electron beam leads to continuous structural changes. Surprisingly, ZZ(57) and ZZ(75) alternatively switched during a period of about 50 seconds (only one exception was observed, and we will explain the reason for this later), with a ZZ(66) transitional state in between, The reversible transition between ZZ(57) and ZZ(75) can be visualized clearly(Figure 2a) and presented in Figure 2g. The HRTEM image series are aligned vertically to demonstrate the edge structural changes(Figure 2a). The heptagons (7 member ring) on the ZZ(57) edges are highlighted, and red and green vertical lines are used to guide the eye. The pentagons at 0s are changed into heptagons at 5s, through the 57-66-75 transition. It should also be noted there are some vacancies left accompanying the reconstructions, i.e. at 0s-4s, one vacancy is created (marked by the red triangle). Moreover, the number of vacancies over the whole edge increases by one during each step of 57-66-75(or 75-66-57) transition. There are two exceptions, from 7s to 10s, the vacancies happen to be refilled by incoming carbon atoms. And from 16s to 29s, a pair of vacancies is added together, and correspondingly, the only 57-66-57 transition case occurs.

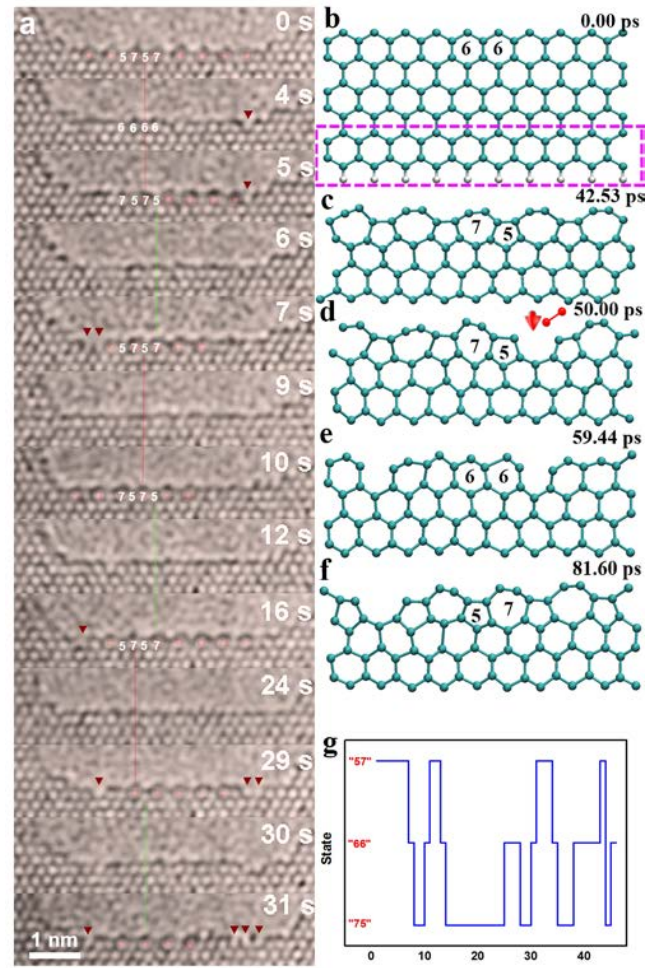

**Figure 2** (a) TEM image series for one graphene zigzag edge transformations during 31s. The heptagons and pentagons are numbered. Red triangles are used to remark the defects created by electron beam. The red and green vertical lines are used to emphasize "5->7" or "7->5" transitions. (b-f) The QM/MD simulations showing the same transformation dynamics as observed in TEM measurements. (g) edge state versus time measured by TEM experiments.

It's intuitively clear that the missing carbon atoms at vacancies are ejected by the knock-on effect of high energy electron beam, considering our experiment are all carried out under room temperature. To better comprehend the relationship between vacancy creation and reconstructions of

ZZ edge, we employed quantum chemical molecular dynamics (QM/MD) simulations based on the density-functional tight-binding (DFTB) potentials.[23] Firstly, displacement threshold energies ($T_d$) for the different carbon atoms on ZZ(57) edge are calculated. $T_d$ refers to the minimal kinetic energy transferred to the nucleus which displaces the atom from recombination with the newly formed vacancy. First, we assume the electron beam knocks on a single C atom from heptagons or pentagons, $T_d$ for each of them are respectively 19.3 and 20.0 eV which already exceed the maximum kinetic energy (16 eV) for one C atom gained from 80 kV beam. Although this mechanism (single vacancy creation) cannot be totally ruled out, we are skeptical about this as we seldom find mono-vacancy resulting in ZZ(-57-56-57-) or ZZ(-57-47-57-) sequences from TEM measurements. When applying velocities to the dimmer carbon of heptagons, $T_d$ significantly decreases to 9.4 eV for each carbon, so there is a tendency that two C atoms in one heptagon have correlated dynamics. Carbon dimmer is knocked from the edge undergoes 5/7 bonds breaking and leaving a free dimmer unit and di-vacancy.[23] Furthermore, the repairing process by a carbon dimmer was also simulated.[23] After the creation of one double atom vacancy(corresponding TEM image in Figure 2a), the ZZ(57) edge is allowed to relax in the MD simulation series (Figure 2b-f). It can be clearly identified that the structural transition develops in the neighborhood of the vacancy. After a short while the reaction front sweeps over the whole edge and a ZZ(66) edge is left, but eventually the ZZ(66) edge is converted into ZZ(75) edge, and all of which is initialized from a single vacancy. The complete transition from ZZ(57) to ZZ(75) consist of two waves, the first is ZZ(57)->ZZ(66) and the second is ZZ(66)->ZZ(75).

Due to the homogeneous illumination in TEM measurement, the shooting of C atoms on ZZ(57)(or ZZ(75)) edges are random. In the triggering step, each loss of carbon dimmer in any of the heptagons (in either ZZ(57) or ZZ(75)) leads to a conversion toward hexagons, and subsequently pentagons. Meanwhile, the newly formed heptagons in the last cycle (from the original pentagon) serve as the source of reconstruction in the next cycle and transform into pentagons. Therefore, the complete reversible transformations on graphene ZZ edges under electron beam influence is: -[ZZ(57)-ZZ(66)-ZZ(75)-ZZ(66)]-. Moreover, if there are more than one dimmer vacancy created in a single step, the order of the transitions can be disturbed and the ZZ(57) state can be restored (29s in Figure 2a). Thus odd number of dimmer vacancies lead to (57)-(66)-(75) transition while even number of double carbon vacancies lead to (57)-(66)-(57) transition. This explains the dependence of transformation order on the odd or even vacancies addition each time (Figure 2a).

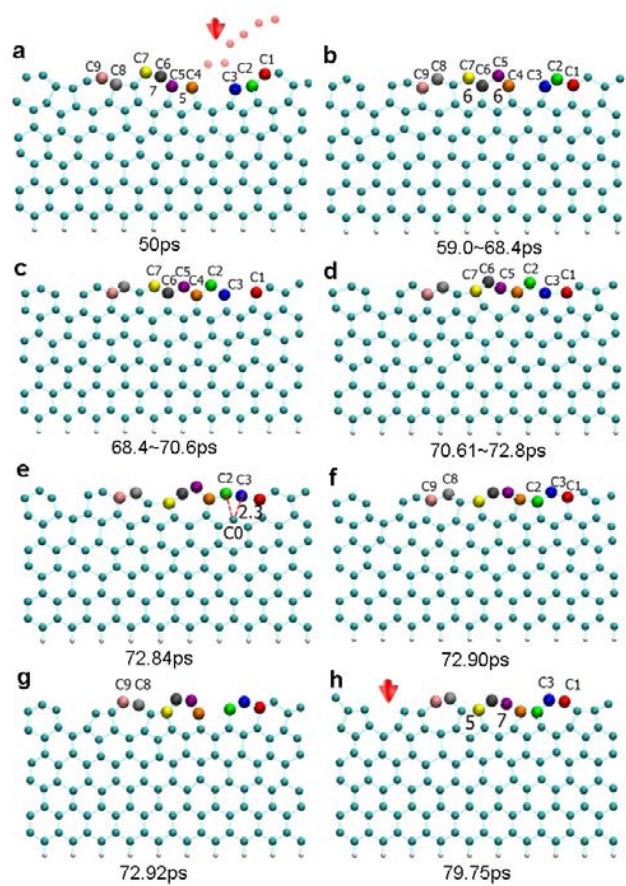

**Figure** 3 (a-h) one ZZ(75) edge in a is changed into ZZ(57) edge in (h) after one defect created (red arrow). The whole process includes two waves starting from the defective point, ZZ(75)->ZZ(66) from (a) to (c) and ZZ(66)->ZZ(57) from (c) to (h). All the related C atoms are marked in colors.

QM/MD allows us getting a deeper insight to understand the rearrangement of (57)-(66)-(75) transition. In Figure 3a, carbon dimmer colored in light red was knocked on leaving a di-vacancy. The long distance of unsaturated C3 and C4 is ca. 3.2 Å which induces edge stress. Directly forming chemical bond between C3 and C4 with two adjacent pentagons is energetically unfavorable, and three pentagons also make the edge bend strongly which is forbidden by 2D graphene backbone. Hence, the carbon edge needs to undergo consequent reconstruction to compensate higher energy induced by the two-atom vacancy. In Figure 3b, we can see series of C2 units rotate toward vacancy. For example, C4-C5 rotates by 30 degree forming a new hexagon within 3ps. C2 unit of C8 and C9 belonging to 57 rings flips by ca. 90 degree, and new vacancy created by 5/5 bond breaking. In Figure 3c, we observe the vacancy can move along the edge without zigzag edge reconstruction because near the vacancy e.g. C2 and C3 of hexagon can freely flip by 120 degree. Di-vacancy results zigzag unstable and further edge rearrangement by C6-C7 shifting to right forming a new pentagon-heptagon in Figure 3d. Note that there is one hexagon on the right of each heptagon, and in Figure 3e we find a transition state (TS) with two equivalent bond lengths of C2-Co and C3-Co of 2.3 Å. Followed by the TS, C4-C2 is formed as another pentagon near the heptagonal ring. From Figure 3f to Figure 3g, C8 and C9 rotating a small degree (ca. 30o) become two vertexes of seven member ring. The left two hexagon pairs rotate the same way as C9 and C8, so finally ZZ(57)-ZZ(66)-ZZ(75) transition ends leaving a shifted position of vacancy. After 100 ps, the structure still remains, and then some 57 rings start to transform into 66 rings resulting 567 mixed rings because of not sufficient large periodic boundary conditions and too high temperature. The MD simulation time we applied are on the scale of nano second (ns), at more than 2000K high temperature to accelerate the dynamics at the edges, while the real time observation in TEM is on the scale of second(s), which is inaccessible with any current simulation methods. The reaction barrier between ZZ(57) and ZZ(66) after triggering is 0.7~0.8 eV by our calculations, a bit higher than the value 0.5eV reported by other groups.[10] Nevertheless, using the thermal activation relationship, $\tau^{-1}=\upsilon_o\exp(-E_b/kT)$, where $\upsilon_o$ is the attempt frequency typically on the THz scale, the escape time τ for ZZ(66)-ZZ(57) reconstructions at room temperature is estimated on the scale of second, in agreement with the time scale of our in situ TEM observations of the transition dynamics. As indicated in our previous report about the Fe atom translocation on graphene edges,[25] the electron beam knock-on effect is the main reason for triggering of the reconstructions(this step is on the timescale of second, confirmed by the C atom shooting rate in TEM), while the following reactions or motion of atoms can be just due to thermal activations (here also on the timescale of second). In TEM, we can count how many carbon atoms are knocked out between each two consequent snapshots, it's unlikely that any reconstructions occur but immediately recover to the original structure between two snapshots that we cannot differentiate, because reconstructions always induce different vacancies at anywhere on the edge, alleviating worries about fast dynamics out of temporal resolution of TEM. Therefore our snapshot rate at 1Hz is sufficient to capture the major dynamics. In addition, the QM/MD simulations can be used to compare with experiments because they have the same reaction barrier and reaction path just under different temperatures.

The edge states of pristine graphene ZZ edge electronic structure have been investigated a lot with theoretical methods like empirical tight binding,[2] DFT,[27] etc. All the studies yield similar results, that the unsaturated π electrons at edges contribute to a dispersiveless band just locating on the Fermi surface which is degenerate starting approximately from the Dirac point in reciprocal space(k=2/3π). The density of edge states decay quickly towards inside of graphene, within 2 to 3 outmost atomic rows.[5] For the reconstructed ZZ(57) edges, calculations were also carried out to show the band structure contains a degenerate edge state near Fermi energy.[10] Therefore, the edge states of both ZZ(57) and ZZ(66) both increase the density of states near Fermi energy, and contribute to the metallic properties. On the other hand, the effect of symmetry on the electrical transport behavior for graphene nanoribbons(GNR) with ZZ(66) edges has been theoretically studied.[28,29] The two subbands near the Fermi level, π and π*, have opposite σ symmetry parity (odd for π and even for π*) when biased by a finite electric field, and the transmission possibility is greatly suppressed. Following the continuous reconstruction on graphene ZZ edges discussed above, our TEM observations uncovered graphene ribbon with different types on the two edges. Figure 4a-c present one ultrathin GNR (five unit cells in width) produced by electron beam irradiation. During 200s, the two edges of this ribbon exhibit ZZ(57), ZZ(66) and ZZ(75) types in a periodic manner, and with different combinations we are able to find three kinds of ZZ GNRs, 57-57, 57-75, and 66-66, as shown in Figure 4a-c, respectively. The geometries of 57-57 and 66-66 GNR have mirror (σ) symmetry while 57-75 not. The pristine 66 edges and reconstructed 57 edges are chemically active owning to the dangling bonds, therefore hydrogen passivation is usually applied.[6] Also to reduce the effect of the electron scattering by σ bonds in transport, in following transport analysis, the graphene edges are all mono-hydrogen passivated.

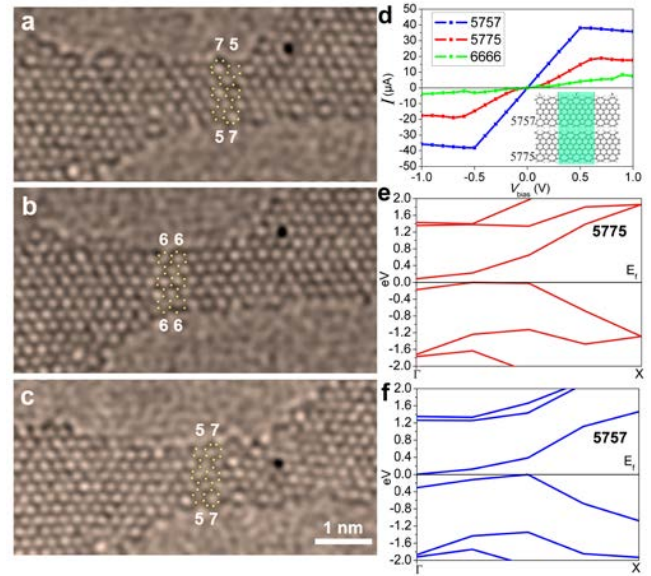

**Figure** 4 (a-c) TEM image series for the same graphene nanoribbon with 7557, 6666, 5757 edges, edges structure modified by electron beam. (d) Quantum transport calculations for the transport property(IV) for the above three kinds of ribbons (same width as in experiments). (e, f) calculated band structures(only close to Fermi Level are shown here) for the 5775 and 5757 graphene nanoribbons which are semiconductor(5775) and metal(5757), respectively.

The ab initio based code for non-equilibrium electrical transport, Transiesta, is utilized to calculate the transport properties of ZZ GNRs.[23] The obtained current versus bias relationship for different types of edge ribbons are shown in Figure 4e. For zigzag edge the current is quite small within 1.0 V bias range (less than 5 μA) although without band gap. It ascribes to zero hopping integral between π and π* electrons with each other with σ symmetry.[28] However, for five-row 57-75 and 57-57 GNRs there is no such restriction of symmetry. U-I dependence of the ultrathin 5775 GNRs behaves as linear within 0~0.52 V, and then reaches saturation. We can find a band gap of 80 meV corresponding to 928 K (from both LDA and GGA predictions) at Γ point. The current of 57-57 GNRs mirror symmetry is the largest one because no band gap can be observed in Figure 4f. Different IV and band structures for three types of edges attribute to their strong edge effects. To verify this, we further calculate the properties of seven row GNRs. The band gap of 5775 disappears and IV relationship is quite similar with 5757's.[23] We notice that because of decreasing edge influence followed by enhanced symmetry effect, the current of 5757 slightly decreases from 40 to 30 μA. The experimental GNRs in Figure 4a-c may not have such big differences as in the transport property calculations, due to the short length of the ribbon and room temperature (calculations in ground state), but the switch effect can be more apparent with improved GNR fabrication method in future. Here our main point is to demonstrate the edge types on opposite graphene zigzag edges can combine and possibly form different GNRs, with distinct properties induced by the edge states.

## Conclusions

To summarize, we have demonstrated the continuous and repeatable 57-66-75-66-57 reconstructions on graphene ZZ edges, by both direct TEM observations (experimental) and DFT, QM/MD simulations. Triggered by the electron beam created defects, this one dimensional phase transition sweep over the whole edge. Hence towards the applications we suggest a new route to control the edge structures of GNRs, say, through defect engineering. This 57-66-75-66-57 reconstructions at ZZ edges make it possible to tune the symmetry of GNRs through controlling the structures of two opposite edges. Finally our quantum transport calculations have demonstrated the metal-insulator-semiconductor transition in the structural transformation of ZZ(57-57)->ZZ(66-66)->ZZ(57-75) GNRs, which can be potentially used for atomic-scale variable electronic devices.

## Methods

### Sample preparation

The original material we used is the commercial HOPG from SPI (ZYA Grade). Using scotch tape we started repeatedly peeling flakes of graphite. Thin flakes left on the tape were released in acetone. Then a TEM copper grid with lacey carbon was dipped in the solution and then washed in deionized water, some flakes can be captured on the grid.

### TEM and multislice image simulation

JEOL 2010F transmission electron microscope fitted with CEOS spherical (Cs) aberration correctors for the objective lens was used. To minimize knock-on damage, the TEM was operated under an acceleration voltage of 80 kV. The samples were measured at room temperature. The electron beam intensity during HRTEM observation is ~0.1pA/nm$^2$. The average background subtraction filtering (ABSF) is carried out on the image post-processing. JEMS software is used for the image simulations. An accelerating voltage of 80kV with an energy spread of 0.3eV, chromatic aberration Cc set to 1mm, spherical aberration Cs set to 1 μm. A defocus of 4nm, and defocus spread of 3nm was implemented. These values are in consistent with the experimental conditions.

### DFT(ab initio) calculation

Density functional theory (DFT) computations were carried out within GGA (PBE)[30,31] as implemented in the DMol$_3$ program package[32,33]. LDA (PWC)[34] is for validated band structures. The basis set was chosen as double numerical-polarized basis set that includes all occupied atomic orbitals with a second set of valence atomic orbitals plus polarized d-valence orbitals. The real-space global orbital cut-off radius was set as 4.8 Å. The interlayer distance was set to 20 Å, which is large enough to minimize artificial interlayer interactions. Fully relaxed geometries were obtained by optimizing all atomic positions until the energy, maximum force, maximum displacement were less than $2\times10^{-5}$ Ha, 0.005 Ha/Å and 0.005 Å, respectively. The k-points samplings were 12×2×1 in the Brillouin zone.

### MD simulation

The MD simulations are performed using the DFTB+ program[35] with self-charge-consistent (SCC) approximation[36] of DFTB potentials in combination with a finite electronic temperature approach (Te = 5000 K). The temperiture is set to 2000 K controlled by Nosé–Hoover thermostat with a coupling constant of 500 cm$^{-1}$.

### Quantum Transport Calculations

The electronic transport properties are conducted by the non-equilibrium Green's function as implemented in the TranSIESTA module within the SIESTA[37].

## AUTHOR INFORMATION

**Corresponding Author**

*zhaojiong@gmail.com

## ACKNOWLEDGMENT

We would like to thank Prof. B. Buchner, Dr. A. A. Popov and IFW Dresden for granting to use TEM facilities. We also would like to thank Prof. You-hua Luo (East China University of Science and Technology) for providing us with the Dmol3 package.

## REFERENCES

1. L.D. Landau, E.M. Lifshitz, Statistical Physics I, Pergamon Press, New York, 1980
2. Nakada, K., Fujita M., Dresselhaus G. & Dresselhaus M.S. Edge state in graphene ribbons: Nanometer size effect and edge shape dependence. *Phys. Rev. B***54**, 17954-17961(1996).
3. Yao, W., Yang, S. A. & Niu, Q. Edge States in Graphene: From Gapped Flat-Band to Gapless Chiral Modes. *Phys. Rev. Lett.***102**, 096801(2009).
4. Yamashiro, A., Shimoi, Y., Harigaya, K. & Wakabayashi, K. Spin- and charge-polarized states in nanographene ribbons with zigzag edges. *Phys. Rev. B***68**, 193410 (2003).
5. Pisani, L., Chan, J. A., Montanari, B. & Harrison, N. M. Electronic structure and magnetic properties of graphitic ribbons. *Phys. Rev. B***75**, 064418 (2007).
6. Wassmann, T., Seitsonen, A. P., Saitta, A. M., Lazzeri, M. & Mauri, F. Structure, stability, edge States, and aromaticity of graphene ribbons. *Phys. Rev. Lett.***101**, 096402 (2008).
7. Song, L. L., Zheng, X. H., Wang, R. L. & Zeng, Z. Dangling bond states, edge magnetism, and edge reconstruction in pristine and B/N-terminated zigzag graphene nanoribbons. *J. Phys. Chem. C***114**, 12145-12150 (2010).
8. Jia X., Hofmann, M., Meunier, V., Sumpter, B. G. *et al.* Controlled formation of sharp zigzag and armchair edges in graphitic nanoribbons. *Science***323**, 1701-1705 (2009).
9. Huang, B., Liu, M., Su, N. Wu, J. *et al.* Quantum manifestations of graphene edge stress and edge instability: a first-principles study. *Phys. Rev. Lett.***102**, 166404 (2009).
10. Rodrigues, J. N. B., Gonçalves, P. A. D., Rodrigues, N. F. G., Ribeiro, R. M. *et al.* Zigzag graphene nanoribbon edge reconstruction with Stone-Wales defects. *Phys. Rev. B***84**, 155435 (2011).
11. Kroes, J.M. H., Akhukov, M.A., Los, J.H., Pineau, N. & Fasolino, A. Mechanism and free-energy barrier of the type-57 reconstruction of the zigzag edge of graphene. *Phys. Rev. B***83**, 165411 (2011).
12. Stone, A. J. & Wales, D. J. Theoretical studies of icosahedral C60 and some related structures. *Chem. Phys. Lett.***128**, 501-503 (1986).
13. Koskinen, P., Malola, S. & Häkkinen, H. Self-passivating edge reconstructions of graphene. *Phys. Rev. Lett.***101**, 115502 (2008).
14. Kotakoski, J., Santos-Cottin, D. & Krasheninnikov, A. V. Stability of graphene edges under electron beam: equilibrium energetic versus dynamic effects, *ACS NANO***6**, 671-676 (2012).
15. Chuvilin, A., Meyer, J. C., Algara-Sillerand, G. & Kaiser, U. From graphene constrictions to single carbon chains. *New Journal of Physics* **11**, 083019 (2009).
16. Koskinen, P., Malola, S. & Häkkinen, H. Evidence for graphene edges beyond zigzag and armchair. Phys. Rev. B **80**, 073401 (2009).
17. Kim, K., Kisielowski, S. C. C., Crommie, M. F., Louie, S.G. *et al.* Atomically perfect torn graphene edges and their reversible reconstruction. *Nat. Comm.***4**, 2723 (2013).
18. Warner, J. H., Lin, Y.-C., He, K., Koshino, M. & Suenaga, K. *Nano Lett.***14**, 6155-6159 (2014).
19. He, K., Robertson, A. W., Fan, Y., Allen, C. S., *et al.* Temperature dependence of the reconstruction of zigzag edges in graphene. *ACS NANO***9**, 4786-4795 (2015).
20. Atkins, P.W., *Physical Chemistry*, W.H. Freeman & Co., New York. (1997).
21. Krasheninnikov, A. V. & Nordlund, K. Ion and electron irradiation-induced effects in nanostructured materials. *J. Appl. Phys.***107**, 071301 (2010).
22. Novoselov, K. S., Geim, A. K., Morozov, S. V., Jiang, D., *et al.* Electric field effect in atomically thin carbon films. *Science***306**, 666-669 (2004).
23. Online supporting materials for this manuscript
24. Zhao, J. *et al.* Free-standing single-atom thick iron membranes suspended in graphene pores. *Science***343**, 1228-1232 (2014).
25. Zhao J. et al. Direct in situ observations of single Fe atom catalytic processes and anomalous diffusion at graphene edges. *Proc. Nat. Acad. Sci.***44**, 15641-15646 (2014).
26. Stronge,W. J. *Impact Mechanics*. Cambridge University Press (2004).
27. Kunstmann, J., Ozdogan, C., Quandt, A. & Fehske, H. Stability of edge states and edge magnetism in graphene nanoribbons, *Phys. Rev. B***83**, 045414 (2011).
28. Li, Z., Qian, H. Wu, J., Gu, B.-L & Duan, W. Role of symmetry in the transport properties of graphene nanoribbons under bias. *Phys. Rev. Lett.***100**, 206802 (2008).
29. Ren Y. & Chen, K.-Q. Effects of symmetry and Stone-Wales defect on spin-dependent electronic transport in zigzag graphene nanoribbons, *J. Appl. Phys.***107**, 044514 (2010).
30. Kohn, W. & Sham, L. J. Self-consistent equations including exchange and correlation effects. *Phys. Rev.* **140**, A1133 (1965).
31. Schluter M. & Sham, L. J. Density functional theory of atoms and molecules. *Phys.Today***2**, 36 (1982).
32. Delley, B. Numerical method for solving the local density functional for polyatomic molecules. *J. Chem. Phys.* 92, 508-517 (1990).
33. Delley, B. From molecules to solids with the DMol$_3$ Approach. *J. Chem. Phys.***113**, 7756-7764 (2000).
34. Perdew J. P. & Yue, W. Accurate and simple density functional for the electronic exchange energy: Generalized gradient approximation. *Phys. Rev. B.***33**, 8800 (1986).
35. Aradi, B., Hourahine, B. & Frauenheim, T. DFTB+, a sparse matrix-based implementation of the DFTB method. *J. Phys. Chem. A***111**, 5678-5684 (2007).
36. Elstner, M., Porezag, D., Jungnickel, G., Elsner, J., Haugk, M., *et al.*Self-consistent-charge density-functional tight-binding method for simulations of complex materials properties. *Phys. Rev. B***58**, 7260 (1998).
37. Soler, J. M., Artacho, E., Gale, J. D., García, A., Junquera, J., *et al.* The SIESTA method for ab initio order-N materials simulation. *J. Phys. Cond. Mat.* **14**, 2745-2779 (2002).